# Inorganic electrochromic metasurface in the visible

*Yohan Lee[1]*, Jonas Herbig[1], Serkan Arslan[1], Dominik Ludescher[1], Monika Ubl[1], Andreas Georg[2], Mario Hentschel[1], and Harald Giessen[1]**

[1] 4th Physics Institute and Research Center SCoPE, University of Stuttgart, Pfaffenwaldring 57, 70569 Stuttgart, Germany

[2] Fraunhofer Institute for Solar Energy Systems, Heidenhofstraße 2, 79110 Freiburg, Germany

*Corresponding author, e-mail: y.lee@pi4.uni-stuttgart.de, giessen@pi4.uni-stuttgart.de

**Keywords:**

Inorganic, Electrochromic, Metasurfaces, Electrically Switchable, Full color switching

**Abstract:**

Color printing based on metallic or dielectric nanostructures has revolutionized color science due to its unprecedented subwavelength resolution. Evidently, the evolution towards the active control of such structural colors with smart materials is in progress for real applications. Here we experimentally demonstrate a large color gamut with high intensity and purity, as well as its switching on and off based solely on tungsten trioxide ($WO_3$) cylindrical resonators. The strong resonances in the visible spectral range in these $WO_3$ metasurfaces can be reversibly switched on and off due to its electrochromism by applying alternating voltages of +2.0 V and -0.3 V. Our approach opens up possibilities for the functional diversification of commercial smart windows, as well as the development of new display technologies in the future.

## Main manuscript

Color generation with nanostructures [1,2], has been explored since ancient times in human history, as seen in the well-known Lycurgus Cup, stained glasses, and potteries [3]. Different from applying dyes or pigments as traditional techniques to generate colors, a structural coloration can offer promising opportunities such as vivid and vibrant colors, high durability, precise color tuning capabilities, and environmental friendliness. Consequently, in the recent decade, there have been extensive efforts employing plasmonic nanostructures towards expanding the color space and achieving high resolution and saturation, because the resonance can be easily tuned by the structural information including the size, shape, and arrangement of the nanoscatteres with recent advancement in nanofabrication and optical characterization techniques [1]. However, due to the inherent loss of metals in the visible range, the resonance peaks from plasmonic nanostructures are typically broad and less intense, preventing them from truly competing with the color space of traditional pigments [4-9]. In this context, high-index dielectric nanostructures such as silicon (Si) or titanium dioxide ($TiO_2$) have garnered significant attention for nanoscale color printing, because they can support optical resonances in the visible spectrum through Mie resonances [10-15]. Such materials possess enhanced resonance quality factors related to the saturation of colors due to the absorption losses.

The notable advantage achievable only through structural color reproduction is the ability to actively control the color via external stimuli, which is not possible with traditional methods. The realization of such dynamic structural color is crucial for extending functionality beyond merely static images to include animations. Commonly reported approaches involve empolying active materials that respond to physical or chemical principles to alter the refractive index or size of nanoresonators [16-24], or to change the refractive index of the background of nanostructures [25-32]. Previous studies on dynamic structural colors have two main limitations: (1) Most studies have focused on transitions to different colors rather than ON/OFF color switching [19-27, 30, 31]. For real applications, it is essential to have several basic colors (such as red, green, and blue) that can be turned on and off, along with their combinations, similar to the principle of commercialized displays. (2) The materials used for color reproduction and the active materials used for color control are different [28, 29, 32]. Adding other types of materials for active control could introduce compatibility issues or compromise the desired optical properties, potentially reducing the overall efficiency or reliability of the color-changing process.

Here, we introduce an electrically switchable full-color palette, made from tungsten trioxide ($WO_3$) nanostructures based on an electrochemically-driven tuning of its refractive index. The proposed platform consisting of solely one material is able to express a wide color gamut, as well as to dynamically switch it on and off via an applied voltage.

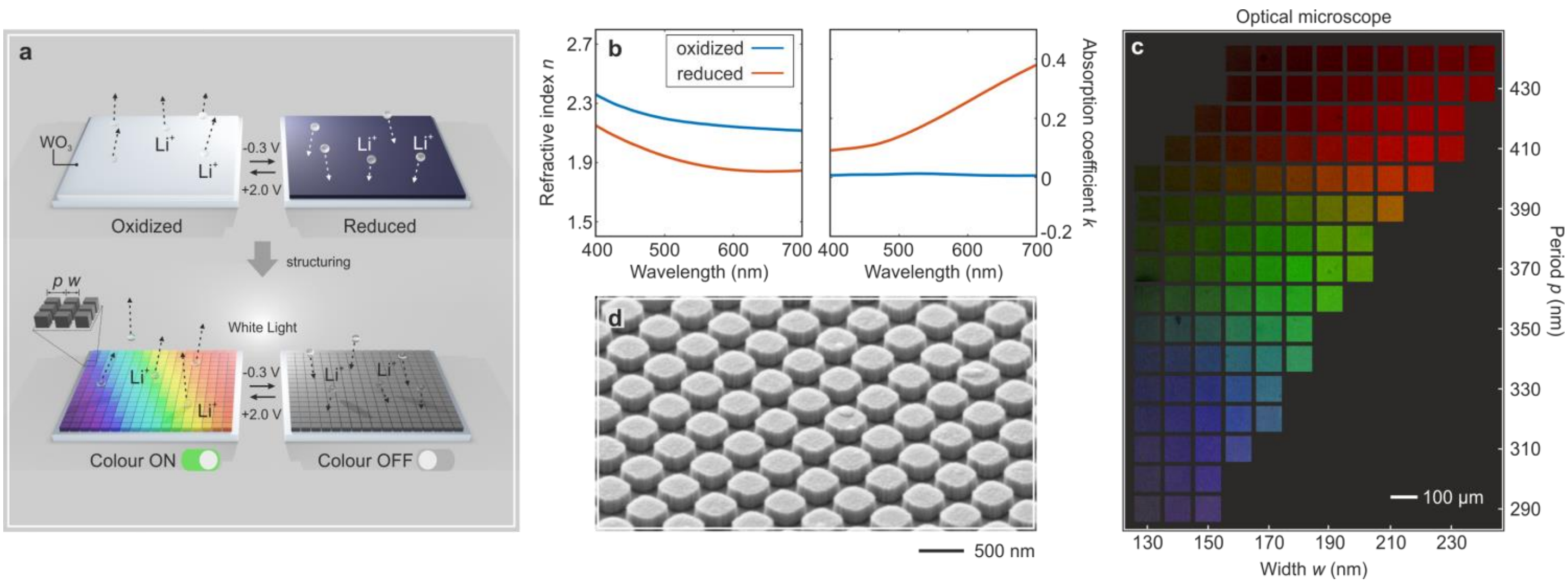

**Figure 1. Concept of full-color-switchable metasurfaces made from tungsten trioxide ($WO_3$).** (a) Schematic of the working principle of the inorganic electrochromic metasurfaces. $WO_3$ as a thin film shows simple electrochromic property from transparent to deep blue with electrochemically redox reaction (Top). Nanopatterned $WO_3$ not only allows for the generation of a wide color gamut depending on the period ($p$) and width ($w$) of nanoresonators, but also allows the colors to be switched on and off at a faster rate (Bottom). (b) Refractive index (Left) and absorption coefficient (Right) of $WO_3$ at oxidized and reduced states. $WO_3$ has relatively high refractive index ($n \approx 2.2$) in the visible. Evidently, nearly no absorption is observed at oxidized state in the visible, while electrochemically reduced $WO_3$ shows dispersive absorption. (c) Optical microscope image with 5x objective of the color palette made from $WO_3$ metasurfaces with the combinations of different $p$ and $w$. (d) Scanning electron microscopy (SEM) image shows the $WO_3$ metasurface with one specific condition of $p$ and $w$.

The concept of our inorganic metasurfaces is illustrated in Figure 1. Pristine $WO_3$, utilized in current smart windows, is transparent in its as-deposited film state (Top of Figure 1a). The dynamic behavior of $WO_3$ is enabled by the unique phenomenon of the material changing color in response to an electric stimulus, known as electrochromism. Depending on the electrochemical reactions induced by the intercalation or de-intercalation of lithium ions dissolved in an electrolyte, it can exhibit a switch from transparency to a deep blue color (see Figure S1 in the Supporting Information). This switching mechanism takes place by the following reaction [33].

$$\mathrm{WO_3} + x\mathrm{Li^+} + x\mathrm{e^-} \rightleftharpoons \mathrm{Li}_x\mathrm{W}_x^{5+}\mathrm{W}_{1-x}^{6+}\mathrm{O_3}$$

In this work, we find that nanopatterned $WO_3$ can express a wide range of color space with high purity by tailoring the geometric conditions of the patterns, such as period $p$ and width $w$ (Bottom of Figure 1a). Furthermore, leveraging the electrochromic phenomenon of the material itself, the brilliant colors can be switched off and on again by applying voltages of -0.3 V and +2.0 V (versus a reference electrode). The proposed concept shows the novelty that color generation and switching are simultaneously achieved solely with monolithic $WO_3$.

The mechanism of our active metasurfaces relies on the variation of refractive index of $WO_3$ by applied voltages. Figure 1b depicts the real and imaginary parts of the refractive indices of $WO_3$ in its oxidized (or pristine) and reduced states (see Table S1 in the supporting Information for detailed values). We notice that in the visible, $WO_3$ has a relatively high refractive index (around 2.2) and exhibits nearly no intrinsic losses when oxidized, enabling it to support high-Q resonance. Although it cannot support strong Mie resonance in a single nanoparticle due to its refractive index being lower than 3.5 in the visible range, unlike silicon or gallium arsenide, it can exhibit lattice resonance by controlling the interactions between the scatterers through periodic arrangement. Notably, as $WO_3$ gets electrochemically reduced by an external voltage, the absorption coefficient of $WO_3$ (the imaginary part of the refractive index) increases, thereby enabling the suppression of such a resonance.

In Figure 1c, we display the color palette generated by our $WO_3$ metasurfaces. Optical images of the colors are taken with a 5x NA 0.15 objective at a fully closed aperture stop, in order to collect only components near normal incidence. (Angle dependency of the metasurfaces is presented in Figure S2 in the Supporting Information.) It is clear that the colors and their intensities are determined by period $p$ and width $w$ of the $WO_3$ nanoresonators. Figure 1d depicts a scanning electron microscope (SEM) image of our $WO_3$ metasurface taken for one of the structures that have different $p$ and $w$. The nanopattern is a cuboid shape as designed, which shows excellent quality of nanofabrication. Additional SEM images of the structures with various conditions can be found in Figure S3 in the Supporting Information. The detail of nanofabrication is provided in Methods and Figure S4 in the Supporting Information.

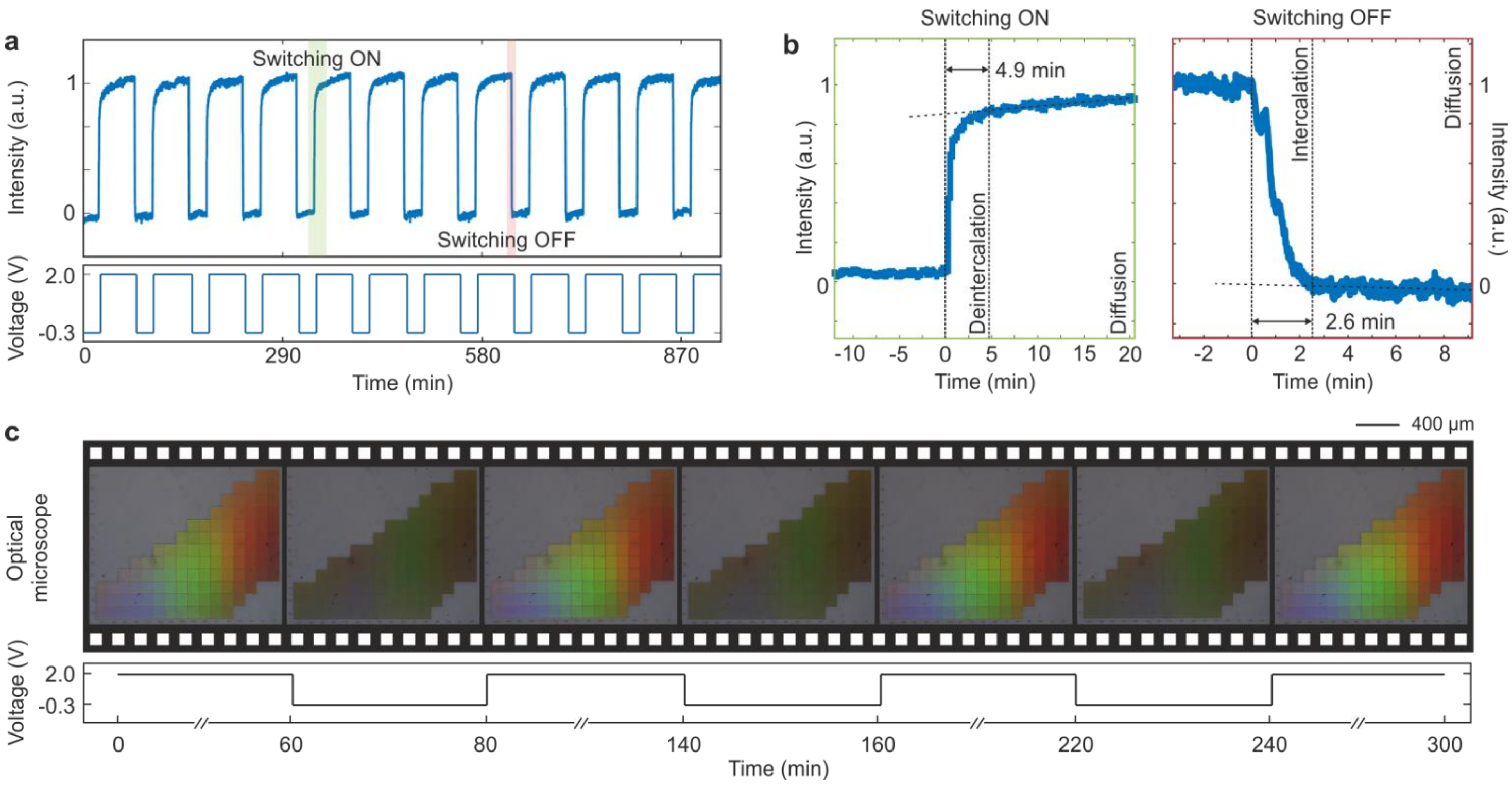

**Figure 2. Switching performance by electrical stimulation of $WO_3$ metasurfaces.** (a) Transmitted intensity through the $WO_3$ metasurface cycling between the ON and OFF states (Top). The green area depicts the switching ON window, while the red area represents the switching OFF window. The voltage range is set between +2.0 V to -0.3 V (Bottom). (b) Plots corresponding to the switching ON and OFF states, as indicated in (a). The cation intercalation/deintercalation and diffusion regions are distinguished based on the point where the rate of change in intensity converges (indicated by the dashed line). The time required for cation deintercalation (Left) and intercalation (Right) is presented. (c) Optical microscope images showing several cycles of colors switching ON and OFF when the applied voltage is cycled in situ between +2.0 V for 60 minutes to -0.3 V for 20 minutes.

The switching performance of our $WO_3$ metasurfaces is investigated in Figure 2. The resonators are switched between ON and OFF states at applied voltages of +2.0 V and -0.3 V while tracking the transmitted intensity (see Methods for the detail). In Figure 2a, the modulated transmitted intensity is depicted in the top graph, corresponding to the voltage applied to the $WO_3$ nanopatterns shown in the bottom graph. When a negative voltage is applied to the electrode, the cations present in the electrolyte are attracted to the $WO_3$ surface. These ions then undergo intercalation into the $WO_3$. This process occurs primarily through electrostatic interactions, where the ions are inserted into specific sites within the $WO_3$, resulting in a rapid reduction of light intensity. After the intercalation, the ions diffuse further into $WO_3$. This diffusion occurs due to the concentration gradient of ions within the $WO_3$, where the ions tend to move towards regions of lower concentration. Similarly, when a voltage is applied in the opposite direction to the electrode, deintercalation occurs first, followed by diffusion, resulting in recovery of light intensity. Twelve switching cycles are illustrated with a period of 80 minutes (20 minutes for OFF and 60 minutes for ON), confirming full switching between ON and OFF states without noticeable degradation. Although further switching is limited by the rapid evaporation of the electrolyte, experimental studies have reported that the switching stability of $WO_3$ is maintained even after more than 100 cycles at the laboratory scale. [34]

Figure 2b displays the analysis of the switching time from ON to OFF (Left) and from OFF to ON (Right). The onset of diffusion is defined as the point at which the rate of change in light intensity becomes constant. The difference in speed between intercalation and diffusion primarily arises from their distinct governing mechanisms. Intercalation occurs rapidly at the electrode-electrolyte interface, driven by charge transfer and electrostatic interactions, whereas diffusion is a bulk process limited by the energy barrier within the electrode lattice. The slower diffusion process results from the need for ions to migrate through the solid structure, which is constrained by concentration gradients and material properties such as ionic diffusivity and lattice openness. We focused on the time required for cation intercalation/deintercalation, as most of the light intensity change occurs during this process rather than the diffusion process. We obtain the deintercalation time of 4.9 minutes and the intercalation time of 2.6 minutes. Intercalation occurs more rapidly as ions are readily inserted into available sites, driven by electrostatic attraction. In contrast, deintercalation is slower due to stronger Coulombic interactions and structural distortions, which increase the energy barrier for ion extraction. It is worth mentioning that the color switching speed during the electrochemical reaction can be affected by several key factors including the thickness of $WO_3$, the cation concentration in an electrolyte, and the substrate temperature. In this work, acetonitrile containing 100 mM $LiClO_4$ as an electrolyte, and room temperature are employed in all of the demonstrations. Acetonitrile is widely utilized in lithium-ion batteries, electrochemical sensors, and high-speed energy storage systems due to its wide electrochemical stability window, high ionic conductivity, and low viscosity. More detailed data on switching speed can be found in Figure S5 in the Supporting Information. Also, the thickness of $WO_3$ is fixed at 200 nm, which directly affects the support of lattice resonance. Details regarding the resonance intensity as a function of thickness can be found in Figure S6 in the Supporting Information.

To visualize the switching performance of our metasurfaces, the images with a 5x objective are depicted in Figure 2c. Furthermore, movies that records the color dynamics of the palette are presented in the Supplementary Movie 1 (for switching OFF) and Movie 2 (for switching ON).

The colors generated by $WO_3$ metasurfaces at a pristine state can be switched off upon lithium ion intercalation induced by an applied voltage of -0.3 V. Importantly, the colors of the palette can be nicely restored with a voltage of +2.0 V. This indicates that the $Li_xWO_3$ are converted back to $WO_3$, therefore, displaying brilliant colors again. Such reversible switching is of high importance for dynamic applications.

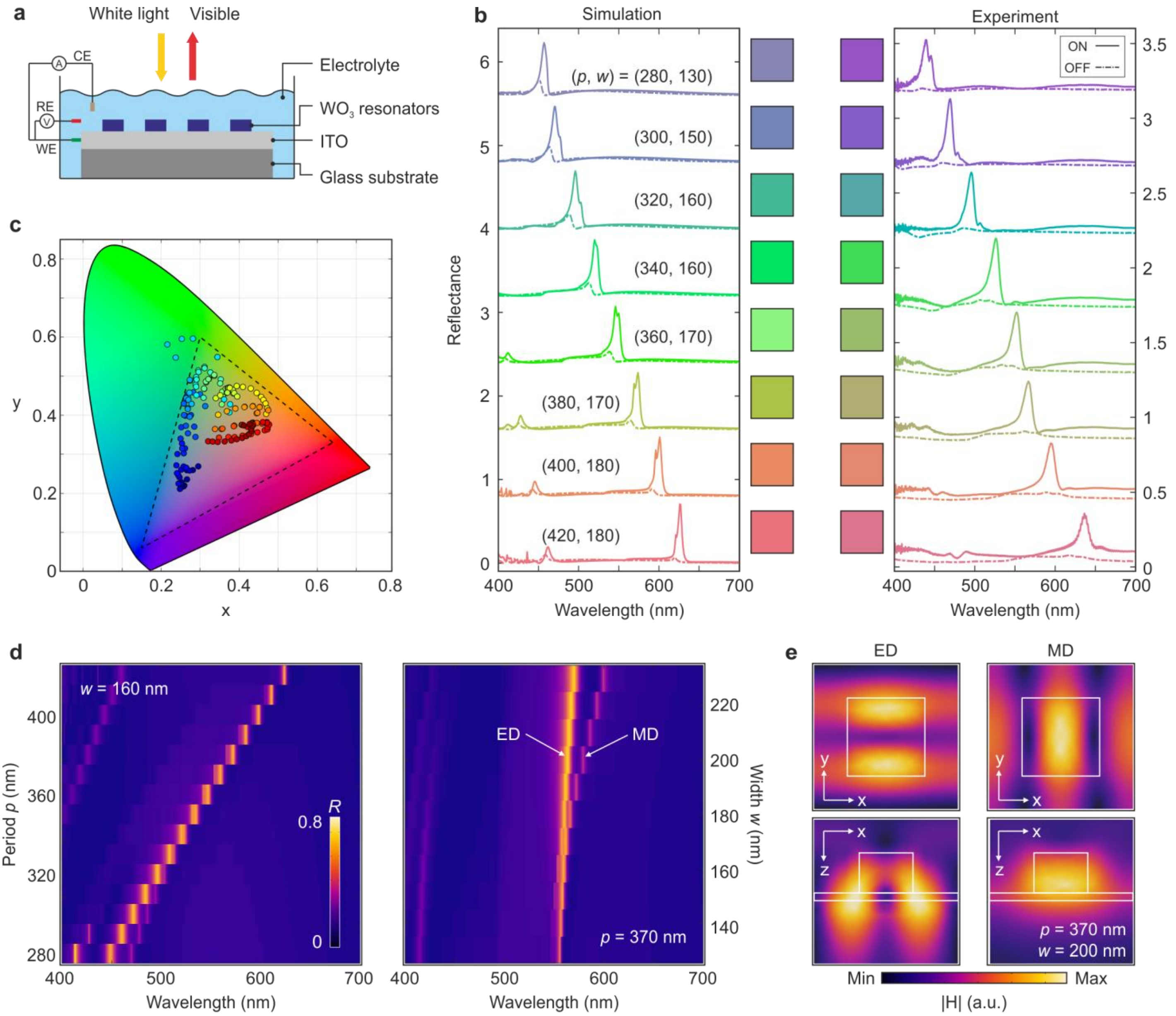

**Figure 3. Optical characterization of the colors from $WO_3$ metasurfaces.** (a) Schematic of the electrochemical cell (three-electrode setup). The $WO_3$ metasurface immersed in the electrolyte is illuminated by a white light source with normal incidence, and the reflected light is captured by an objective. (b) Simulated and experimental normal reflectance spectra for ON and OFF states. For the ON state, the colors corresponding to the reflectance spectra are depicted in the adjacent squares. (c) Measured reflectance spectra plotted in CIE 1931 color map. A large color gamut is obtained by varying the parameters of the metasurfaces (period *p* and width *w*). The triangle formed by the dashed lines in the plot represents the standard RGB (sRGB) color space. (d) Simulated reflectance spectra as a function of wavelength for different periods or widths. (Left) With the width fixed to 160nm, high reflectance with narrow bandwidth is mainly depending on the period of nanostructures. (Right) Electric dipole (ED) and magnetic dipole (MD) resonances are observed depending on the widths with the period fixed to 370 nm. (e) The magnetic field

distributions of the cross-sections at the ED and MD resonances, respectively, in $WO_3$ resonators with $p$ = 370 nm and $w$ = 200 nm.

Color ON/OFF switching via external voltages is carried out in a liquid electrolyte using an electrochemical cell with a three-electrode system (Figure 3a; see Methods). For both ON and OFF cases, the spectral analysis of the representative colors among various structural information (period $p$ and width $w$) is carried out and presented in Figure 3b. The simulated spectra (left panel) indicate that the reflectance resonance has a high peak (around 70 %) with a narrow bandwidth, resulting in a high purity color. Notably, the remaining reflectance which is slightly blue-shifted can be seen even in OFF state. It can be visually observed during the color change in Figure 2c (the colors are not completely 'OFF', and become slightly different). This is because our mechanism of ON/OFF switching is based on electrochromism of $WO_3$, which is the absorption tuning by electrical stimulus. The absorption coefficient of $WO_3$ increased by electrochemical reduction is not sufficiently high to suppress the entire reflected intensity. The experimental reflectance spectra and their colors (Right panel) show an overall good agreement with respect to the simulation results. From the measured reflectance spectra, the structural colors that can be expressed through the combination of $p$ and $w$ of $WO_3$ metasurfaces are displayed in the CIE 1931 color space (Figure 3c).

To elucidate the underlying physics of the observed spectral characteristics, we numerically calculated the reflectance while varying $p$ and $w$, respectively (Figure 3d). Above all, the reflectance peak clearly exhibits red-shifts with increasing period $p$ (left panel), meaning the colors from our metasurfaces are mainly based on lattice resonance [35]. Also, we can observe the typical characteristics where the dipole resonance shifts according to the structural parameter $w$ of the scatterers (right panel). Figure 3e depicts magnetic field distribution of the cross-section at the resonant peaks with $p$ = 370 nm and $w$ = 200 nm. From Figure 3e, we infer that the mode at shorter wavelength is the electric mode, whereas the long wavelength mode is the magnetic resonance. In principle, each $WO_3$ nano-cuboid can support a Mie resonance due to its relatively high refractive index. However, its refractive index is not sufficiently high to be excited in a single resonator like silicon. In this sense, we arrange $WO_3$ scatterers into periodic metasurfaces to enhance the reflectance efficiency due to the coupling effect between dipole resonances, and to suppress the other high-order resonant modes as a photonic band gap [13, 36].

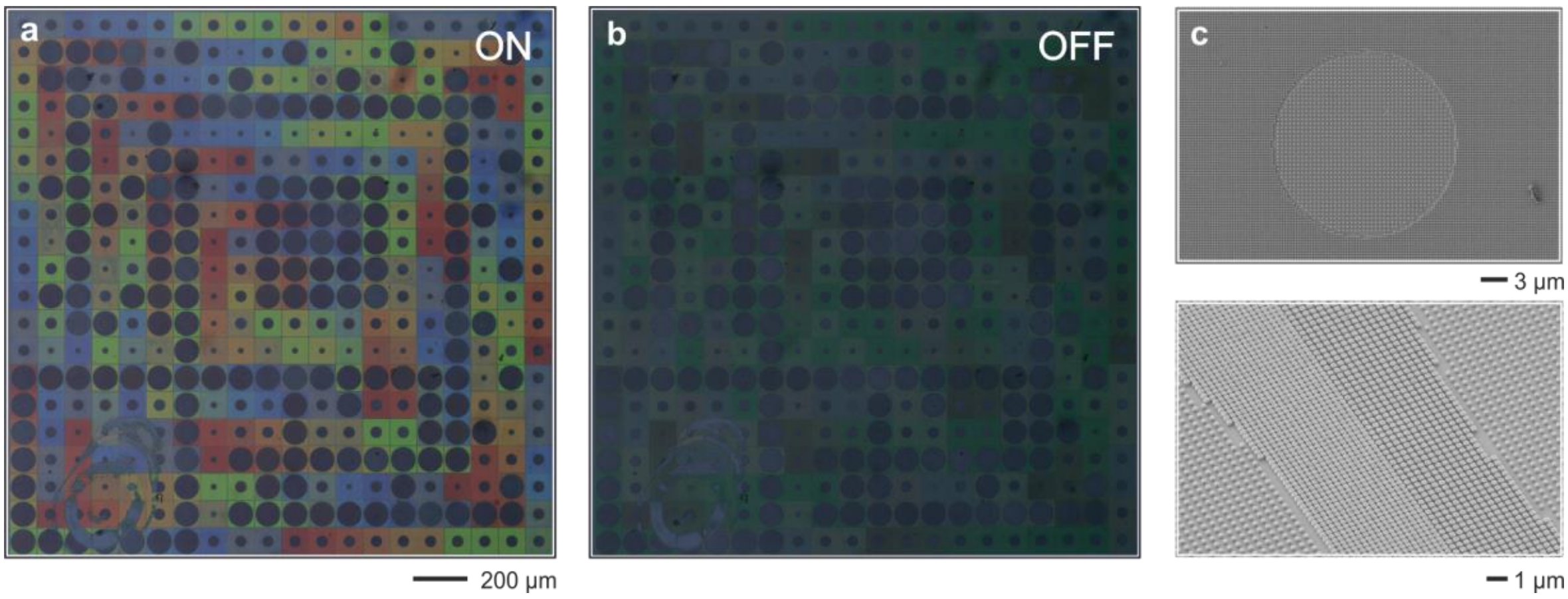

**Figure 4. Dynamic color switching of arbitrary image.** (a) Optical microscopic image of the modified artwork, inspired by 'Orion Gris' of Victor Vasarely. Nanopatterned $WO_3$ is ON state in electrochemically oxidized state. The size of whole pattern is 2 mm by 2 mm. (b) The artwork in OFF state during electrochemically reduced reaction for color switching. (c) Enlarged SEM images of fabricated sample (Top: one square building block including a circle, Bottom: the edge between building blocks)

$WO_3$ metasurfaces have the potential to go beyond the light and thermal energy blocking capabilities of conventional smart windows [37], enabling the reproduction of arbitrary color images and their ON/OFF switching. To demonstrate this ability, we proposed a millimeter-sized artwork as a blueprint inspired by Victor Vasarely's *Orion Gris* (Figure 4a). The basic structure is a square shape containing a circle in the center. Each structure has circles of different sizes and backgrounds of different colors. Figure 4b represents the image erased into the OFF state for a dynamic switching. The SEM images of the basic structure and the edge area are shown in Figure 4c. It is worth mentioning that those SEM images were taken from the sample that already got switched on and off several times. This means that $WO_3$ nanopatterns still exhibit no noticeable degradation during lithium ion intercalation and de-intercalation within a safe voltage range.

In summary, we have introduced novel active metasurfaces that operate in the visible spectrum. The high refractive index and electrochromic property of $WO_3$ enable a high purity color gamut with unprecedented subwavelength resolution and dynamic ON/OFF switching via applied voltages. Structural engineering and optimization of electrochemical environment can be carried out to further improve the switching speed and larger color space for real world applications. Our concept will be of importance for solid-state devices with electrically tunable optical responses and reflective displays, especially once the engineering task of addressing single pixels has been solved.

## Methods

### Structure fabrication and characterization

As substrate we use ITO (50 nm) coated glass to operate the ON/OFF switching of reflected colors and allow electrical addressability. A 200 nm layer of $WO_3$ is deposited onto the substrate using reactive DC-magnetron sputtering (Star 100-TetraCo) with a metallic tungsten target. The deposition is performed with a power of 0.58 kW, a pressure of $1.1\times10^{-2}$ mbar, and an Ar flow rate of 200 sccm, and an $O_2$ flow rate of 70 sccm. The substrate was not heated and its temperature during deposition was below 100°C. The film is covered with a double-layered poly(methyl-methacrylate) (PMMA) positive tone resist (Allresist AR-P 642.06 200k, Allresist AR-P 672.02 950k), used to define nanostructures with electron beam lithography (EBL). After development in methylisobutylketone (MIBK), we use electron-gun evaporation to deposit a 30 nm Chromium (Cr) layer. After lift-off, Cr serves as hard-mask for the subsequent chemical etching process, which removes uncovered $WO_3$. The $WO_3$ film is etched in a PlasmaPro80 etcher from Oxford with 100 W of radio-frequency power with an atmosphere of 10 sccm of $N_2$, 15 sccm of $CF_4$, and 10 sccm of $SF_6$. As a result, we obtain a $WO_3$ metasurface after removing Cr via a commercial Cr remover. Scanning electron microscope images are taken with a Zeiss SEM Gemini 560.

### Spectral measurements

The presented reflectance spectra were measured in a setup based on an upright optical microscope (Nikon Eclipse LV). The exit path of the microscope is connected to the entrance slit of a Princeton Instruments grating spectrometer (Iso plane 160) with a Peltier-cooled CCD camera (PIXIS 256). The sample is placed on a motorized XY stage (Märzhäuser) and illuminated through the objective (Nikon TU Plan ELWD 20x, NA 0.4) by the in-built halogen white light lamp. We ultilized internal apertures of the microscope to restrict the image plane (field stop) to the 200 µm by 200 µm $WO_3$ metasurface arrays. Additionally, the aperture in the microscope Fourier plane (aperture stop) is closed to restrict incident and collected light to angles close to normal incidence and blocking light scattered from the surrounding substrate. The measured spectra are normalized to the reflectance of a protected aluminum mirror (Thorlabs).

**Numerical simulations**

Numerical simulations were carried out using commercial software COMSOL Multiphysics based on a finite element method. Periodic boundary conditions were used for calculation of the structure arrays. The refractive index of $SiO_2$ was taken as 1.47.

**Electrochemical switching of metasurfaces**

We utilize a custom-built electrochemical cell to electrically switch the $WO_3$ metasurfaces using a three-electrode configuration. The cell is sealed with thin glass window at the top and bottom to allow optical access during reflectance/transmittance experiments. It includes side ports for electrolyte inflow and outflow, as well as connections for the reference and counter electrode. ITO (for electrical contact) underneath the $WO_3$ nanopatterns serves directly as the working electrode. The counter electrode is a platinum wire, and the reference electrode is silver/silver-choride (Ag/AgCl), both in contact with 100 mM $LiClO_4$ electrolyte. A potentiostat (BioLogic SP-200) regulates the voltage over time. Temporal and switching characteristics are analyzed using a 633 nm He-Ne laser and a photodiode.

**Author contributions**

Y.L conceived the original idea. Y.L., M.H. and H.G. supervised the work. Y.L. performed simulations and modelling for the $WO_3$ nanopatterns. A.G. provided $WO_3$ films. Y.L., M.U., and M.H. fabricated and characterized the samples. Y.L., J.H., and D.L. measured the color samples. S.A. performed the spectroscopic measurements. All authors participated in the preparation and writing of the manuscript.

**Acknowledgements**

This work was funded by Alexander von Humboldt Foundation (Y.L.), Bundesministerium für Bildung und Forschung (H.G.), Deutsche Forschungsgemeinschaft (GRK2642 Photonic Quantum Engineers; H.G.), European Research Council (ERC Advanced Grant Complexplas & ERC PoC Grant 3DPrintedOptics; H.G.), Carl-Zeiss-Stiftung (Center Qphoton, EndoPrint3D; H.G.). Also, this research was supported by Basic Science Research Program through the National Research Foundation of Korea (NRF) funded by the Ministry of Education (2021R1A6A3A14043838; Y.L.).